\documentclass[twocolumn]{revtex4}
\usepackage{graphicx,amsmath,amssymb,mathrsfs}
\usepackage{epsfig}

\begin{document}

\title{Testing axion physics in a Josephson junction environment}

\author{Christian Beck}

\affiliation{Queen Mary University of London, School of Mathematical Sciences, Mile End Road, London E1 4NS, UK}

%\pacs{85.25.Cp} {Josephson devices}
%\pacs{14.80.Va} {Axions}
%\pacs{85.25.Dq} {SQUIDS}

\begin{abstract}
We suggest that experiments based on Josephson junctions, SQUIDS,
and coupled Josephson qubits may be used to construct a resonant
environment for dark matter axions.
%have a cosmological interpretation in terms
%of axionic dark matter physics, in the sense that they allow for
%What do advanced
%Josephson junction technologies,
%SQUIDs, coupled Josephson qubits
%and related superconducting devices used in nanotechnology
%have in common with
%the problem of dark matter in the early universe? A lot more than
%might seem obvious at first sight, as will be shown in this
%letter. One of the major candidates for dark matter in the
%universe is the axion.
%The equation of motion of the axion
%misalignment angle and that of the phase difference in a Josephson junction are identical
%if the symbols in the mathematical equations are properly re-interpreted.
%analogue simulation of early-universe axion physics.
%For incoming present-day
%axions such a Josephson junction environment may act as a resonant medium.
%Our approach for the first time
%connects two so far unrelated scientific fields and opens up novel
%prospects for axionic dark matter detection.
We propose
experimental setups in which
axionic interaction strengths
in a Josephson junction environment can be tested, similar in nature to
%It also allows to develop new experimental setupe that constrain axionic
%interaction processes in a Josephson environment,
%(and the cosmological re-interpretation of already existing Josephson junction experiments)
%that constrain axionic dark matter interaction strengths in a Josephson junction environment,
recent experiments
that test for quantum entanglement of two coupled Josephson qubits.
%It allows also for the development of new types of
%detectors that search for phase synchronization effects.
We point out that the parameter values relevant for early-universe
axion cosmology are accessible with present day's achievements in
nanotechnology. We work out how typical dark matter and dark energy signals
would look like in a novel detector that exploits this effect.
\end{abstract}

%\keywords{xxx}
\maketitle

\section{Introduction}

Research in nanotechnology is currently at a very advanced stage, and so is
research in
cosmology. The two scientific fields
are proceeding independently of each other,
and the two scientific
communities don't know each other --- dealing apparently with very different
subject areas.
But are these two research areas really that far apart? At first glimpse, certainly they
are.
But a look at the equations of motions of Josephson junctions,
SQUIDS (superconducting quantum interference devices), coupled Josephson qubits and similar
superconducting devices used in nanotechnology \cite{tinkham,josephson,koch,steffen,blackburn}
on the one hand and axionic dark matter \cite{tegmark,sikivie2,duffy}
on the other indicates that it makes sense to think
about common approaches in both areas. The equations of motion
 are very much the same (with a suitable
re-interpretation of the symbols used) and hence it makes sense
to
%develop common approaches and
translate known results from one area (nanotechnology) into
possible results and phenomena for the other area (axion cosmology).
%This will be worked out
%in the following.

There are many different candidates for dark matter in the universe,
with extensive experimental searches,
WIMPS (weakly interacting massive particles)
\cite{bernabei, CDMS2, aalseth} and axions
\cite{tegmark, sikivie2, duffy} being the most popular ones.
In contrast to WIMPS, axions are very
light particles that behave similar to a cold quantum gas \cite{sikivie2}.
Axions have been around as models for dark matter for quite a while, and
there are several
experiments that search for them directly in the laboratory,
using e.g.\ cavities and strong magnetic fields, which trigger the decay of axions into
two microwave photons \cite{sikivie2, duffy, cavity}.
These microwave photons are in principle detectable if the cavity
resonates with the axion mass
(see e.g. \cite{duffy} for a formula for the
expected power generated by axion-photon conversions). Searches have been unsuccessful so far, but it is necessary to
scan a huge spectrum of cavity frequencies, for which SQUID amplifiers with very low noise levels are a very
useful technological tool to reduce the noise level
and to improve the scanning efficiency
\cite{cavity}.
%\footnote{These applications of SQUIDs
%for axion searches are very important from an efficiency point of view
%for cavity experiments, but different
%from the fundamental physics analogies between axions and Josephson junctions that we want to emphasize
%in the following.}.
Quasi-axionic particles play also an important role in
topological insulators, new materials with exotic metallic states on
their surfaces \cite{moore,wilczek,qi}. These recent developments
illustrate that it does make sense
to look at axion physics in a much broader
context than within
the original
model, which was
concerned with the solution of the strong CP problem in the
standard model of elementary particle physics \cite{peccei}.

The axion is characterized by a phase angle,
the axion misalignment angle \cite{tegmark,sikivie2,duffy}.
There are different types of axion production mechanisms
in the early universe,
topological axion production and vacuum alignement (see e.g. \cite{duffy}
for more details).
 On the other hand, a Josephson junction is
also characterized by a phase, namely the phase difference of the macroscopic
wave function describing the two superconducting electrodes of the junction.
As will be shown in this letter,
the equation of motion of the axion misalignment angle is identical to that
of the phase difference of a resistively shunted Josephson junction, with a suitable re-interpretation
of the currents involved. This opens up the theoretical
possibility to connect both fields, and to make analogue experiments
simulating
axion cosmology using superconducting devices in the laboratory.
Moreover, this novel approach also opens up the possibility to test for interaction strengths
of incoming present-day axionic dark matter in a given resonant Josephson junction environment.
Suggested future experiments of this type can be easily performed in the laboratory and
%A major result of this letter is the
%prediction of new phenomena
%for axionic dark matter (based on known Josephson junction
%physics) which can be experimentally tested in the laboratory
%and which
may ultimately open up the route for novel
methods of dark matter and dark energy detection based on modern nanotechnology.

\section{Equations of motion of the axion and of Josephson junctions}

%Let's startcompare the mathematics underlying both axions and Josephson junctions.
Consider an axion field $a=f_a \theta$, where $\theta$
is the axion misalignment angle and $f_a$ is
the axion coupling constant. We consider such an axion in the early
universe, described by a Robertson Walker metric, and neglect spatial
gradients.
The equation of motion of the axion misalignment angle $\theta$
is
\begin{equation}
\ddot{\theta} +3H \dot{\theta}+ \frac{m_a^2c^4}{\hbar^2} \sin \theta = 0. \label{e1}
\end{equation}
Here $H$ is the Hubble parameter and $m_a$ denotes the axion mass. The forcing term $\sin \theta$
is produced by QCD instanton effects. In a mechanical analogue, the
above equation is that of a pendulum in a constant gravitational field
with some friction determined by $H$.

If strong external electric and magnetic fields
$\vec{E}$ and $\vec{B}$ are present, then the axion couples as follows:
\begin{equation}
\ddot{\theta} +3H \dot{\theta}+ \frac{m_a^2c^4}{\hbar^2} \sin \theta = \frac{g_\gamma}{\pi}
\frac{1}{f_a^2} c^3 e^2 \vec{E} \vec{B} \label{1}.
\end{equation}
$g_\gamma$ is a model-dependent dimensionless coupling constant describing the decay of the axion
into two photons ($g_\gamma =-0.97$ for KSVZ axions (Kim-Shifman-Vainshtein-Zakharov axions \cite{ksvz1,ksvz2}),
$g_\gamma=0.36$ for DFSZ axions (Dine-Fischler-Srednicki-Zhitnitsky axions \cite{dfsz1,dfsz2})).
The typical parameter ranges that are allowed for dark matter axions are
\cite{sikivie2, duffy}
\begin{equation}
6 \cdot 10^{-6}eV \leq m_ac^2 \leq 2 \cdot 10^{-3} eV \label{333}
\end{equation}
and
\begin{equation}
3 \cdot 10^{18} eV \leq f_a \leq 10^{21} eV.
\end{equation}
The
product $m_ac^2f_a$ is expected to be of the order $m_ac^2 f_a \sim 6 \cdot 10^{15} (eV)^2$.

Let us now compare this with the equations of motion of
resistively shunted Josephson junctions (RSJs).
A Josephson junction consists of
two superconductors with a thin insulator inbetween \cite{josephson}.
Such a Josephson junction can be a very complex system, with all
kinds of interesting effects, in particular for small junctions.
However, in the simplest theoretical models
the phase difference $\delta$
of the macroscopic wave function of the two superconductors satisfies
\begin{equation}
\ddot \delta +\frac{1}{RC} \dot{\delta} +\frac{2eI_c}{\hbar C}
\sin \delta = 0 \label{e2} .
\end{equation}
Here $R$ is the shunt resistance, $C$ is the capacity of the junction, and $I_c$ is the critical
current of the junction.
The frequency
\begin{equation}
\omega =\sqrt{\frac{2eI_c}{\hbar C}}
\end{equation}
is called the
plasma frequency of the Josephson junction. The product
\begin{equation}
Q:= \omega RC
\end{equation}
is the so-called quality factor of the junction.

If a bias current $I$ is applied to the junction by maintaining
a voltage difference $V$ between the two superconducting electrodes, then the equation of
motion becomes
\begin{equation}
\ddot \delta +\frac{1}{RC} \dot{\delta} +\frac{2eI_c}{\hbar C} \sin \delta = \frac{2e}{\hbar C} I. \label{2}
\end{equation}
Remarkably, the equations of motions of axions and of
RSJs are identical provided we make the following identifications
in eqs.~(\ref{1}) and (\ref{2}):
\begin{eqnarray}
3H&=& \frac{1}{RC} \label{3} \\
\frac{m_a^2c^4}{\hbar} &=& \frac{2eI_c}{C} \label{4} \\
\frac{g_\gamma}{\pi f_a^2} c^3 e^2 \vec{E} \vec{B} &=& \frac{2e}{\hbar C} I. \label{5}
\end{eqnarray}

Let us now further work out this interesting connection.
We note that it is possible to make analogue experiments
with RSJs that
simulate axion cosmology in the laboratory. To simulate an axion
in a certain era of cosmological evolution, one builds
up an RSJ which has its parameters $R,C, I_c, I$ fixed by eqs.~(\ref{3})-(\ref{5}).
The left-hand side are cosmological parameters, the right-hand side is
nanotechnological engineering.
Essentially the inverse Hubble constant $H^{-1}$ (the age of the universe) fixes the product $RC$,
the axion mass fixes the critical current $I_c$ and the axion coupling to external electromagnetic
fields $\vec{E} \vec{B}/f_a^2$ is represented by the
strength of the bias current $I$.

\section{Relevant parameter values}

It is remarkable that the numerical values of the parameters for
typical axionic dark matter physics
and for typical Josephson junction experiments have similar order of magnitude.
Let us start to illustrate this with some old experiments that were ground-breaking
technology some 30 years ago, the Josephson junction experiments performed
by Koch, Van Harlingen and Clarke in \cite{koch}.
They built up four different samples
of Josephson junctions
with parameters values in the range
$R \sim 0.075-0.77 \Omega$, $C \sim 0.5-0.81 pF$, $I_c \sim 0.32-1.53 mA$.
According to eqs.~(\ref{3})-(\ref{5}), these experiments of Koch et al.
thus simulate the dynamics of
axion-like particles in a very early universe whose age is
of the order $H^{-1}=3RC \sim 10^{-13}-10^{-12}$
seconds and where the axion mass
is in the range
$0.98 \cdot 10^{-3}-1.58\cdot 10^{-3}$ eV.
This simulated axion mass is just at the upper end of what is
expected for dark matter axions, see eq.~(\ref{333}).

By increasing either the shunt resistor
$R$ or the capacity $C$ to larger values, one can simulate axion cosmology at a later time,
for example during or after the QCD phase transition time
($H^{-1}\sim 10^{-8}s$)
where axion physics becomes most relevant.
Current nanotechnology covers this range.
For example,
the recent experiment by Steffen et al. \cite{steffen},
which tests for quantum entanglement of coupled Josephson qubits,
is classically well described by
an RSJ model where the
product $3RC$ is of the order
$H^{-1}=3RC=1.2 \cdot 10^{-6}s$ \cite{blackburn}. This experiment
thus simulates the
dynamics of weakly coupled axions
{\em after} the QCD phase transition in the early universe.
%Hence in our analogy
%QCD phase transition, and hence at a time where axionic dark matter cosmology is
%quite relevant, the axion having obtained a non-zero mass.

It is quite interesting to look at recent
experiments with Josephson junctions
and to check which axion masses
these correspond to, based on our identification in eq.~(\ref{4}).
The experiment of Steffen et al. \cite{steffen}
corresponds
to an axion mass of $m_ac^2=\hbar \omega = 3.3 \cdot 10^{-5}eV$, much smaller
than for the Koch et al. experiment \cite{koch} but again within the range expected
for dark matter axions, see eq.~(\ref{333}).
%It is interesting to see that recent experiments
%\cite{penttilae,nagel,steffen, blackburn,takahide}
%simulate axions with masses in the entire range of what is interesting
%from a dark matter point of view.
The experiments of Penttillae et al.
\cite{penttillae}, dealing with superconductor-insulator
phase transitions, simulate $m_ac^2=1.32 \cdot 10^{-4}eV$.
Nagel et al. \cite{nagel} report on negative absolute resistance effects
in Josephson junctions, these experiments formally have
$m_ac^2=2.83 \cdot 10^{-5}eV$.
Superconducting atomic contacts \cite{della-rocca} correspond to even smaller
axion masses, namely $m_a=6.7 \cdot 10^{-6}eV$,
at the lower end of what is allowed in eq.~(\ref{333}).
Two-dimensional Josephson arrays,
as built up in \cite{takahide},
correspond to arrays of coupled axions with
$m_ac^2$ in the range $6.62 \cdot 10^{-5}-1.52\cdot 10^{-4}eV$.
Remarkably, all these recent experiments are within
the range of axion masses that are of interest from a
dark matter point of view.
They can thus be regarded as simulating axionic physics with realistic
parameters.
%They can thus be regarded as realistic
%analogue experiments simulating an axionic dark matter environment
%in the early universe.
The main point of this letter, which we will work out in more detail in the following,
is that a given Josephson junction environment may indeed serve for incoming cosmological
axions as a kind of resonant medium. This opens up new possibilities of axionic dark
matter detection.

\section{Possible coupling mechanism of axions to Josephson junctions}

Inspired by the above qualitative and quantitative agreement of the relevant
equations of motion,
we may consider the possible existence
of interaction phenomena for axionic dark matter that are inspired by
known effects in Josephson junctions.
For example, it is well-known that two Josephson
junctions may couple in a SQUID-like structure. Can axions form a similar
SQUID-like state? Moreover, there is the Josephson effect, of utmost
importance in many technological applications. This effect does not only exist
for superconducting devices connected by a weak link but also for
Bose Einstein condensates (BEC) \cite{levy}.
Given that axionic dark matter behaves similar to a BEC \cite{sikivie2},
it is natural to ask
whether the analogue of the Josephson effect exists for axionic dark matter.
The mathematical equations allow for such an effect.
%Many other effects, such as e.g. Shapiro steps
%that occur if Josephson junctions are irradiated
%with external radiation, may occur for axions
%in an analogous way.

Let us first work out the theoretical possibility that axions
are able to form SQUID-like structures.
We just sketch
the main idea.
%Consider an RSJ which has a plasma frequency $\omega_p=\sqrt{\frac{2eI_c}{\hbar C}}$ close to
%the axion mass $m_a$, according to eq.~(\ref{4}), and that is driven by
%an external bias current $I>I_c$. Free axion quanta correspond to
%small nearly-harmonic oscillations of the misalignment angle $\theta$,
%in accordance with eq.~(\ref{e1}).
%Suppose such an axion enters a Josephson junction that has similar parameters
%as the entering axion.
%Then this basically means we
%have a system of two Josephson junctions, the second one represented by the entering axion.
It is well-known that if two Josephson junctions,
the first one having phase difference $\delta$ and the second one having phase difference
$\theta$, are put together in a SQUID-like configuration, then
the two phases $\delta$ and $\theta$ start to synchronize, according to the general
equation
%\begin{figure}
%\includegraphics[width=8cm]{axion-fig1.eps}
%\caption{Two Josephson junctions forming a SQUID. If there is a small magnetic
%flux $\Phi$ through the SQUID, then the two phase differences will
%synchronize.}
%\end{figure}
\begin{equation}
\delta -\theta=2\pi \frac{\Phi}{\Phi_0} \mod 2\pi. \label{flux}
\end{equation}
Here $\Phi$ is the magnetic flux enclosed by the SQUID, and $\Phi_0=\frac{h}{2e}$
is the flux quantum. Eq.~(\ref{flux}) is a simple consequence of the fact that the gauge-invariant
phase of the SQUID must be single-valued. As a matter of fact, if the flux $\Phi$ enclosed
by the SQUID is
much smaller than $\Phi_0$, then the above condition
implies
\begin{equation}
\delta = \theta,
\end{equation}
i.e. both phases are synchronized.
A mechanical analogue would be that of two pendula in a gravitational field,
with masses of similar order of magnitude (in our case represented by the Josephson
plasma frequency). The pendula will ultimately
synchronize their movements. We would expect that axions can form
similar synchronized states. One could in fact think of entire clumps of
dark matter axions
(in analogy to arrays of Josephson junctions, well-known in
nanotechnology) which are coupled and perform synchronized motion. As a whole,
they look like a spatially extended dark matter particle of bigger mass.

A related interesting theoretical idea would be that axions could
weakly couple to ordinary Josephson junctions
if these possess similar parameter values. This would be of utmost interest
for detection purposes.
Classically, the coupling between two Josephson junctions with
shunt resistance $R$, capacity $C$, and inductivity $L$ is
described by the following coupled differential equations
\cite{blackburn}:
\begin{eqnarray}
\ddot \delta +\frac{1}{RC} \dot \delta +\omega^2 \sin \delta &=&\gamma_x (\ddot \theta -\ddot \delta) +\frac{1}{CL}
(\delta +2\pi M_1) \nonumber \\
\ddot \theta +\frac{1}{RC} \dot \theta +\omega^2 \sin \theta &=&\gamma_x (\ddot \delta -\ddot \theta) +\frac{1}{CL}
(\theta +2\pi M_2) \nonumber \\
\, & \, & \, \label{coupled}
\end{eqnarray}
$\delta$ is the phase difference of the first junction, $\theta$ that of the
second junction.
$M_i=\Phi_i/\Phi_0$ is the normalized flux
enclosed by junction $i$ ($i=1,2$), and $\gamma_x=C_x/C$ is a
small dimensionless coupling constant, assuming both
junctions are capacitively coupled by a capacity $C_x$.
For example, in the experiments of Steffen et al.
dealing with coupled Josephson qubits \cite{steffen}, $\gamma_x = 2.3 \cdot 10^{-3}$.
Usually the
damping term proportional to $\dot \delta$ and $\dot \theta$ is neglected in
the theoretical treatment of these types of experiments \cite{blackburn}.
%The above classical equations of motion describe
%quite well the experimentally observed
%phenomena \cite{steffen, blackburn}.

In our case, if a similar effect is to be exploited for
dark matter detection purposes, then
the phase $\delta$ would describe
an ordinary Josephson junction and the phase $\theta$
an axion that passes through this Josephson junction.
%The term $\frac{1}{CL}
%(\theta +2\pi M_2)$
%describes a small additional effective self-interaction potential of
%not required it can be made zero by letting $L \to \infty$.
One of the simplest coupling schemes that could be experimentally
tested would
given by
\begin{eqnarray}
\ddot \delta +\frac{1}{RC} \dot \delta +\omega^2 \sin \delta &=&\gamma_x (\ddot \theta -\ddot \delta)
\nonumber \\
\ddot \theta +3H \dot \theta +\frac{m_a^2c^4}{\hbar^2} \sin \theta &=&\gamma_x (\ddot \delta -\ddot \theta) .
\label{coupled2}
\end{eqnarray}
This corresponds to the case $L\to \infty$ in eq.~(\ref{coupled}).
For present-day axions, it certainly
makes
sense to neglect the friction term $3H$, just as
the corresponding friction term was neglected in the theoretical treatment of \cite{blackburn}
for two coupled Josephson qubits.
Formally,
present-day axions have a very high quality factor
$Q= \omega RC = m_ac^2/(3\hbar H)$, because the universe is very old.

If the axion mass is at resonance with the Josephson
plasma frequency, $m_ac^2=\hbar \omega$,
then synchronization effects of the phases $\delta$ and $\theta$ will occur
if $\gamma_x$ is not too small,
just as they occur for coupled Josephson qubits \cite{steffen,
blackburn}.
One could also allow for an axion coupling to fluxes similar as in eq.~(\ref{coupled}), in this case
the axion would have a small additional self-interaction potential
given by
$V(a)=-\frac{1}{CL}(\frac{1}{2}a^2+2\pi M_2 f_a a)$.
%Of course, other coupling schemes than the simple form (\ref{coupled2})
%could be considered as well.
Quantum mechanically, one could even
speculate on the formation of
entangled states between axions and Josephson qubits.
But of course we are still far away from such novel types of detectors at the moment.

In fact, nothing is known on the size of the dimensionless coupling
$\gamma_x$ describing the coupling of an axion to a
given Josephson junction environment.
While the trivial solution
$\gamma_x =0$ is certainly possible, $\gamma_x >0$ is not forbidden
by any first principle. Given the quantitative agreement between
the parameters of axion physics and Josephson junction physics outlined
in section 3, one should consider the possibility
that $\gamma_x$ might again be of similar order of magnitude
as in current nanotechnological experiments. This can be experimentally tested.

There are no astronomical
constraints on the size of $\gamma_x$
 since almost all of the matter in the universe is not
in the form of Josephson junctions. Hence the only way to constrain $\gamma_x$
is to scan a range of plasma frequencies and look for the
possible occurence or non-occurence
of universal resonance effects, produced by axions of the dark matter halo
that hit terrestrial Josephson junction experiments.
The intensity of this effect should display small yearly modulations, just similar as in
the DAMA/LIBRA experiments \cite{bernabei}.
 What corresponds to tuning the cavity
frequency in the experiments \cite{cavity} would correspond to tuning
the plasma frequency $\omega$ in these new types of
nanotechnological dark matter experiments.

It has been previously suggested \cite{beck1,beck2,beck3} that Josephson junctions could also be
used as a laboratory test ground for dark energy.
The basic idea here is as follows:
The fluctuation dissipation theorem (FDT) \cite{kubo,beck4} predicts
in addition to deterministic bias currents $I$ the existence
of noise currents $I_N$ with a universal power spectrum given by
\begin{equation}
S(\nu)=\frac{4}{R} \left(\frac{1}{2} h \nu+ \frac{h \nu}{e^{\frac{h \nu}{kT}}+1}\right) \label{noise}
\end{equation}
on the right-hand side of eq.~(\ref{2}). But the FDT is very general, it also
predicts noise currents with the same power spectrum
on the right-hand side of eq.~(\ref{1}) and (\ref{coupled2}).
Following the line
of arguments presented in \cite{beck1, beck2, beck3} the linear
term $\frac{1}{2}h\nu$ could
be influenced by zeropoint fluctuations of a dark energy field,
whatever its nature, that is coupled into the macroscopic quantum system via
a phase synchronization mechanism \cite{beck3}. For example, this could be
vacuum fluctuations associated with an axion field, thus connecting dark energy
and dark matter.
A simple toy
model would be that the dark energy vacuum fluctuations kick the
Cooper pairs via phase synchronization effects while they tunnel. Whereas ordinary electromagnetic
zeropoint fluctuations exist at any frequency, vacuum fluctuations associated with this
dark energy field would only exist below a critical frequency
in the THz region because the
dark energy density of the universe is finite. If this hypothesis is true then a cutoff
of the
measured noise power spectrum
at a critical frequency corresponding to dark energy density should exist,
which can be experimentally tested \cite{beck3}.
Clearly, from a new physics point of view, it is very important
to test the validity of eq.~(\ref{noise}) at frequencies corresponding to
the dark energy scale, similarly as it is very important to test
gravitational forces at these scales
\cite{adelberger} or the dependence of the Casimir effect on a
postulated cutoff \cite{peri}. In this way Josephson junctions could
provide suitable experimental setups for tests on both dark energy
and (axionic) dark matter.

\section{Axionic Josephson effect}

Given the similarity between axions and Josephson junctions,
we may also consider
the existence of an axionic Josephson effect, similar in spirit
to what was experimentally observed for BEC in \cite{levy}.
An RSJ biased with voltage $V$ generates Josephson radiation with frequency
\begin{equation}
\hbar \omega_J =2eV.
\end{equation}
For such a biased junction the phase $\delta$ grows linearly in time, i.e.
\begin{equation}
\delta (t) = \delta
 (0) +\frac{2eV}{\hbar} t
\end{equation}
and the relation between bias current $I$ and applied voltage $V$ is
\begin{equation}
V= R \sqrt{I^2-I_c^2} \approx RI \mbox{$\;\;\;$ for $I>>I_c$}.
\end{equation}
Josephson oscillations set in if
\begin{equation}
I>I_c, \label{ic}
\end{equation}
i.e. the bias current $I$ must be larger than the critical current $I_c$
of the junction. In the mechanical analogue, the pendulum rotates
with large kinetic energy.
%\begin{figure}
%\includegraphics[width=4cm]{inverted-pendulum.eps}
%\caption{In the mechanical analogue, the Josephson effect corresponds
%to a pendulum that rotates with large kinetic energy.}
%\end{figure}

According to eq.~(\ref{1}) and (\ref{2}), the axion
misalignment angle $\theta$ also starts to increase linearly in time if it is being
forced by very strong products of $\vec{E}$ and $ \vec{B}$ fields. So from
a formal mathematical point of view, an
axionic Josephson effect is possible. We get
\begin{equation}
\hbar \omega_J =2eV \approx 2e RI = \frac{g_\gamma}{\pi} \frac{1}{f_a^2} c^3 \frac{e^2}{3H }\hbar \vec{E} \cdot
\vec{B}, \label{11}
\end{equation}
where in the last step we used eq.~(\ref{3}) and (\ref{5}).
Condition (\ref{ic}) translates to
\begin{equation}
\frac{g_\gamma}{\pi} c^3 e^2 \hbar^2 \vec{E} \vec{B} > f_a^2 m_a^2 c^4 =: \Lambda^4 .  \label{ic2}
\end{equation}
QCD-inspired models of axions require $\Lambda \approx 78 $ MeV \cite{tegmark}.
One can easily check
that the strength of the $\vec{E}$- and $\vec{B}$-fields required to observe axionic Josephson oscillations
must be much higher than anything that can be produced in the laboratory.
%Strong magnetic fields in the laboratory correspond to about $10$
%Tesla, and strong electric fields
%to about $10^9V/m$. This gives $\vec{E} \cdot \vec{B} \approx 10^{10}VT/m$. Typical masses and coupling parameters
%of the axion that are considered in the literature would require the
%product $\vec{E} \cdot \vec{B}$ to be
%extremely large for this effect to set in, roughly larger by a factor
%$10^{30}$ of what can be easily produced in the laboratory.
%However, the axion mass $m$
%is known to be temperature-dependent and in the very early universe,
%where the temperature is very high, the axion mass $m$ it is close to zero.
%Under these circumstance the condition
%(\ref{ic2}) is less stringent, thus allowing in principle for the
%existence of axionic Josephson oscillations.
%However, in astrophysical sources such as
%merging neutron stars, magnetic fields can be extremely strong, of sufficient size for the
%above condition to be satisfied \cite{price}.
%Moreover, in the very early universe the (temperature-dependent)
%axion mass is much smaller \cite{tegmark} and hence
%inequality (\ref{ic2})
%is easier to satisfy.

However, allowing the possibility of phase synchronization,
due to a $\gamma_x \not= 0$ in eq.~(\ref{coupled2}), there is
another interesting possibility here.
If we assume that phase synchronization
sets in for some axions hitting a Josephson junction with resonant
plasma frequency,
then a moderate bias current $I$ in the Josephson
junction environment would simulate for these synchronized axions
the formal existence a huge
product $\vec{E} \cdot \vec{B}$ according to eqs.~(\ref{1}), (\ref{2}), (\ref{5}). The huge
magnetic field formally seen by these synchronized axions
will make the
axions decay into two microwave photons, which could then be detected,
for example in form of Shapiro steps (Shapiro steps are well-known step-like structures in
the $I-V$ curves of irradiated Josephson junctions \cite{shapiro}).
Again this theoretical idea opens up the possibility to develop
new detectors for
axionic dark matter
\footnote{
A strong resonance of unknown origin, observed in the experiments of
Koch et al.\cite{koch} at 368 GHz for the 4th junction
of their experimental series would point to an axion
mass of $1.52 \cdot 10^{-3}$ eV if interpreted in this way.}.
%If this (speculative) interpretation of the measured data is correct, then the plasma
%frequencies of these junctions would indeed be close to the axion mass and forming
%a resonant medium for incoming dark matter axions. Further experimental tests and
%calculations are however necessary to confirm or refute such
%an interpretation.

If the voltage $V$ is applied to a Josephson junction whose superconducting electrodes
are separated by a distance $d$, then this generates the electric field strength
$E=V/d$ between the electrodes. Synchronized axions would oscillate with
the same Josephson frequency as the Josephson junction. They would thus formally
see the magnetic field
\begin{equation}
B= \frac{6\pi f_a^2 d H}{g_\gamma c^3 \hbar e} \label{juelich}
\end{equation}
in the same direction as $\vec{E}$.
This expression is obtained by putting $E=V/d$ in eq.~(\ref{11}).
The result is independent of the applied voltage $V$.
Note that although
today the Hubble parameter $H$ is very small $(H\sim 2\cdot 10^{-18}$ $s^{-1}$), eq.~(\ref{juelich})
formally generates large magnetic fields
(for $d\sim 10^{-6}m$ we get $B\sim 10^5T$). These large (virtual) magnetic fields immediately
make the synchronised axion decay into microwave photons.

\section{Suggestion for experimental searches}

Let us now discuss how a future experiment to detect dark matter axions
with Josephson junctions could look like.
We consider a detector consisting of many coupled Josephson junctions.
Assume a dark matter axion
with mass $m_a$ enters the tunneling region of one of the Josephson junctions.
All
Josephson junction
are driven by bias currents $I>I_c$. If the plasma frequency is at resonance with the
axion mass, then with a certain probability phase synchronization between the
axion phase $\theta$ and the Josephson phase $\delta$ will set in.
The axion will then
formally see a huge magnetic field $\vec{B}$ given by eq.~(\ref{juelich}), because for it
the Josephson environment of the detector looks like a set of fellow axions driven by
the left-hand side of eq.~(\ref{5}).
This huge magnetic field does not exist in reality, it is just briefly
seen by the axion as long its phase is synchronized with the resonant Josephson
environment. The huge (virtual) magnetic field will make the axion decay into
two microwave photons. These microwave photons can then be detected.

Microwave photons produced by axion decays will
modify the $I-V$ curve of the Josephson junction, by producing
Shapiro steps.
The theory of irradiated Josephson junction
(Josephson junctions subject to external electromagnetic radiation)
is well-established \cite{tinkham}. One
knows that in the $I-V$ curve
of the Josephson junction discontinuities appear, at fractions of frequencies
of the incoming photons. These are the so-called Shapiro steps \cite{shapiro}. The
theory is well-established and we refer to the textbooks \cite{tinkham}. The new things in our case
would be that
Shapiro steps produced by axion decays should now also occur if no external radiation is applied,
since photons are internally created by axion decay.
We may call this 'spontaneous' Shapiro steps.
The observation of spontaneous Shapiro steps would be a clear signature of axions.
To exploit this effect,
the detector should be carefully shielded against any external electromagnetic
radiation.

Let us give a rough estimate of signal strengths.
From astronomical
observations, the dark matter halo density of galaxies is known to be about
$\rho_{halo}=0.17...1.7 \cdot 10^{6}$ $GeV/m^3$.
Suppose a large part of this is due to
axions, i.e.\ $\rho_a \sim \rho_{halo}$.
The number density $n$ of axions is then given by $n=\rho_a /m_ac^2$.
Suppose the earth moves through a cloud of axions with relative velocity $\vec{v}$.
Then the flux of
axions through our detector (number of axions $N_a$ per time unit $T$) is given by
\begin{equation}
\frac{N_a}{T}=n v A= \frac{\rho_a vA}{m_ac^2}
\end{equation}
where $A$ denotes the total area of tunneling regions
of all Josephson junctions perpendicular to $\vec{v}$.
With a certain probability $\eta$ a given axion
crossing the detector will resonate with the
Josephson junction environment and
decay into detectable microwave photons.
The number $\eta$ is unknown and but can be
experimentally constrained in future experiments.
The total power produced by decaying axions is
\begin{equation}
P=\eta \frac{N_a m_ac^2}{T}=\eta  \rho_a vA.
\end{equation}
Suppose the detector is large, say $A\sim 1m^2$, and that
$v$ is given by the movement of the sun relative to the
galactic centre,
\begin{equation}
v=(2.3\pm0.3)\cdot 10^5 \frac{m}{s} \approx 10^{-3}c.
\end{equation}
The term $\pm 0.3$ describes yearly modulations \cite{bernabei},
$v$ has a maximum in June
and a minimum in December.
Then the dissipated power
due to decaying axions is
of the order $P\sim \eta 10 W$. This is detectable even for
very small values of $\eta$.
If the dissipated power is really due to axions from the galactic halo,
then
a yearly modulation of about 10\%, similar as in the DAMA-LIBRA experiment \cite{bernabei},
with a maximum in June and a minimum in December should be observed. All this can be easily
checked in future experiments.

\section{Conclusion}

Whatever the conditions were in the very early universe, Josephson junctions, SQUIDS,
and spatially extended arrays of these superconducting devices can nowadays be built
for a wide range of different parameters $R,C,I_c$, and it is very easy to tune
the bias current $I$ to any value of interest.
It is also very easy to adjust the plasma frequency of a Josephson
junction to any value of interest.
It is thus possible to simulate an
axionic dark matter environment in the early universe by
building up the corresponding Josephson junctions with parameters
given by eqs.~(\ref{3})-(\ref{5}).
In particular, it is possible to investigate more realistic models of
spatially-temporally coupled axions, by building up arrays of
Josephson junctions. If the coupling $\gamma_x$ is different from 0, then these
Josephson junctions can form a resonant environment for incoming dark matter
axions.

There is the prospect of
developing new generations of detectors for dark matter axions that
search for possible resonance effects and phase synchronization
if the Josephson plasma frequency is close to the axion mass.
In this way the size of the coupling $\gamma_x$ between
axions and a given Josephson environment could be
experimentally constrained. Noise measurements could also
be made to constrain certain types of dark energy models \cite{beck3}.
%We suggest to
%systematically scan the entire
%range of possible axion masses given by eq.~(\ref{333}).
As outlined above, the
relevant dark matter and dark energy parameter range is accessible by modern technological
developments in nanotechnology.
An obvious advantage of these types of experiments is that
the formal existence of extremely large products of electric and magnetic field strengths
$\vec{E} \vec{B}$ acting on the axion can be simulated by a very simple
experimental setup, an easily tunable bias current $I$, assuming that
some axions hitting the Josephson junction will synchronize their phase
due to a non-zero $\gamma_x$.
This effect may be exploited in the future
to develop new types of axionic dark matter and dark energy detectors based
on modern nanotechnology.


\begin{thebibliography}{99}
\bibitem{tinkham} M. Tinkham, {\em Introduction to Superconductivity}, Dover (2004)
\bibitem{josephson} B.D. Josephson, Phys. Lett. {\bf 1}, 251 (1962)
\bibitem{koch} R.H. Koch, D. Van Harlingen, J. Clarke,
Phys. Rev. B {\bf 26}, 74 (1982)
\bibitem{steffen} M. Steffen et al., Science {\bf 313}, 1423 (2006)
\bibitem{blackburn} J.A. Blackburn, J.E. Marchese, M. Cirillo, N.
Groenbech-Jensen, Phys. Rev. B {\bf 79}, 054516 (2009)
\bibitem{tegmark} M.P. Hertzberg, M. Tegmark, F. Wilczek,
Phys. Rev. D {\bf 78}, 083507 (2008)
\bibitem{sikivie2} P. Sikivie, Q. Yang, Phys. Rev. Lett. {\bf 103},
111301 (2009)
\bibitem{duffy} L.D. Duffy, K. van Bibber, New J. Phys. {\bf 11},
105008 (2009)
\bibitem{bernabei} R. Bernabei et al., Eur. Phys. J. C {\bf 56},
333 (2008)
\bibitem{CDMS2} CDMS II collaboration, Science {\bf 327}, 1619 (2010)
\bibitem{aalseth} C.E. Aalseth et al., Phys. Rev. Lett. {\bf 106}, 131301 (2011)
%\bibitem{sikivie1} P. Sikivie, Phys. Rev. Lett. {\bf 51}, 1415 (1983)
\bibitem{cavity} S.J. Asztalos et al., Phys. Rev. Lett {\bf 104}, 041301 (2010)
\bibitem{moore} J.E. Moore, Nature {\bf 464}, 194 (2010)
\bibitem{wilczek} F. Wilczek, Nature {\bf 458}, 129 (2009)
\bibitem{qi} X.-L. Qi et al., Phys. Rev. B {\bf 78}, 195424 (2008)
\bibitem{peccei} R.D. Peccei and H. Quinn, Phys. Rev. Lett. {\bf 38},
1440 (1977)
\bibitem{ksvz1} J.E. Kim, Phys. Rev. Lett. {\bf 43}, 103 (1979)
\bibitem{ksvz2} M.A. Shifman, A.I. Vainshtein, V.I. Zakharov, Nucl. Phys. B {\bf 166}, 493 (1980)
\bibitem{dfsz1} M. Dine, W. Fischler, M. Srednicki, Phys. Lett. B {\bf 104},199 (1981)
\bibitem{dfsz2} A.R. Zhitnitsky, Sov. J. Nucl. Phys. {\bf 31}, 260 (1980)
\bibitem{penttillae} J.S. Penttil\"a et al., Phys. Rev. Lett. {\bf 82},
1004 (1999)
\bibitem{nagel} J. Nagel et al., Phys. Rev. Lett. {\bf 100}, 217001 (2008)
\bibitem{della-rocca} M.L. Della Rocca et al., Phys. Rev. Lett. {\bf 99},
127005 (2007)
\bibitem{takahide} Y. Takahide et al., Phys. Rev. Lett. {\bf 85}, 1974
(2000)
\bibitem{beck1} C. Beck, M.C. Mackey, Phys. Lett. B {\bf 605}, 295 (2005)
\bibitem{beck2} C. Beck, M.C. Mackey, Physica A {\bf 379}, 101 (2007)
\bibitem{beck3} C. Beck, M.C. Mackey, Int. J. Mod. Phys. D {\bf 17}, 71 (2008)
\bibitem{kubo} R. Kubo, Rep. Prog. Phys. {\bf 29}, 255 (1966)
\bibitem{beck4} C. Beck, M.C. Mackey, Fluct. Noise Lett. {\bf 7}, C31 (2007)
\bibitem{adelberger} D.J. Kapner et al., Phys. Rev. Lett. {\bf 98}, 021101
(2007)
\bibitem{peri} L. Perivolaropoulos, Phys. Rev. D {\bf 77}, 107301 (2008)
\bibitem{levy} S. Levy, E. Lahoud, I. Shomroni, J. Steinhauer,
Nature {\bf 449}, 579 (2007)
%\bibitem{price} D.J. Price, S. Ross, Science {\bf 5}, vol 312, No 5774, p. 719 (2006)
\bibitem{shapiro} S. Shapiro, Phys. Rev. Lett. {\bf 11}, 80 (1963)
\end{thebibliography}
\end{document}